\documentclass[10pt,journal,twocolumn]{IEEEtran}
\IEEEoverridecommandlockouts
\usepackage{cite}
\usepackage{amsmath,amssymb,amsfonts,booktabs}
\usepackage{graphicx}
\usepackage{textcomp}
\usepackage{xcolor}
\usepackage{enumerate}
\usepackage{epstopdf}
\usepackage{subcaption}
\usepackage{amsthm}

\newtheorem{remark}{Remark}

\usepackage{enumitem}

\newtheorem{proposition}{Proposition}

\usepackage{algorithm}
\usepackage{algorithmic}

\allowdisplaybreaks[4]

\def\BibTeX{{\rm B\kern-.05em{\sc i\kern-.025em b}\kern-.08em
		T\kern-.1667em\lower.7ex\hbox{E}\kern-.125emX}}

\begin{document}

\title{Full-Duplex Integrated Sensing, Communication, and Computation over Low-Altitude\\Wireless Networks}

\author{
        Yiyang Chen,
        Wenchao Liu,
        Xuhui Zhang,
        Jinke Ren,
        Huijun Xing,
        Shuqiang Wang,
        Yanyan Shen,
        and Kim-Fung Tsang

\thanks{
Yiyang Chen is with Southern University of Science and Technology, Guangdong 518055, China, and also with Shenzhen Institutes of Advanced Technology, Chinese Academy of Sciences, Guangdong 518055, China (e-mail: yy.chen1@siat.ac.cn).
}

\thanks{
Wenchao Liu is with the School of Automation and Intelligent Manufacturing, Southern University of Science and Technology, Shenzhen 518055, China (e-mail: wc.liu@foxmail.com).
}

\thanks{
Xuhui Zhang and Jinke Ren are with Shenzhen Future Network of Intelligence Institute, the School of Science and Engineering, and the Guangdong Provincial Key Laboratory of Future Networks of Intelligence, The Chinese University of Hong Kong, Shenzhen, Guangdong 518172, China (e-mail: xu.hui.zhang@foxmail.com; jinkeren@cuhk.edu.cn).
}

\thanks{
Huijun Xing is with the Department of Electrical and Electronic Engineering, Imperial College London, London SW7 2AZ, The United Kingdom (e-mail: huijunxing@link.cuhk.edu.cn).
}

\thanks{
Shuqiang Wang, Yanyan Shen, and Kim-Fung Tsang are with Shenzhen Institutes of Advanced Technology, Chinese Academy of Sciences, Guangdong 518055, China (e-mail: yy.shen@siat.ac.cn; sq.wang@siat.ac.cn; kftsang@ieee.org).
}

}

\maketitle

\begin{abstract}

With low-altitude economies emerging as a pivotal sector, this study explores an integrated sensing, communication, and computation system over low-altitude wireless networks. A full-duplex autonomous aerial vehicle (AAV) operates as an AAV-enabled low-altitude platform (ALAP), concurrently executing data transmission, target sensing, and mobile edge computing services. To minimize systemic energy consumption under sensing beampattern constraints and computational demands, we formulate an optimization problem coordinating task allocation, computation resource allocation, and transmit/receive beamforming. Given the non-convexity and highly variable coupling, an efficient iterative convex approximation framework based on alternating optimization decomposes the problem into tractable subproblems. Moreover, the convergence and computational complexity of the proposed algorithm are rigorously analyzed. Simulations verify up to $54.12\%$ energy savings versus benchmarks.
\end{abstract}


\section{Introduction}
\IEEEPARstart{T}{he} low-altitude economy (LAE), an emerging cyber-physical-economic paradigm enabled by low-altitude wireless networks, has emerged as a critical research frontier across both the academia and industrial community in the next-generation smart city infrastructures and autonomous aerial systems development \cite{10681882, liu2025movable}.
Specifically, the LAE operates within a three-dimensional (3D) heterogeneous airspace over the low-altitude wireless networks, encompassing multi-layer economic activities such as ground-based intelligent manufacturing, autonomous aerial operations, and dynamic infrastructure provisioning.
At its core, low-altitude platform (LAP) forms the operational backbone of the LAE, referring to various aircraft operating within the low-altitude airspace, along with their associated ground support systems and service frameworks \cite{10693833}.
While the LAE applications demonstrate pivotal potential across intelligent logistics networks, agricultural pest control, and real-time emergency response systems, establishing robust air-ground connectivity between the LAPs and heterogeneous ground users remains a significant challenge.
To ensure reliable and efficient communication, autonomous aerial vehicle (AAV), also known as unmanned aerial vehicles or drones, enabled LAP (ALAP) has become a cornerstone strategy in modern low-altitude wireless networks, driven by their unique advantages in establishing line-of-sight (LoS) channels, dynamic 3D mobility, and cost-effective scalable network deployment through automated configuration and intelligent resource orchestration \cite{8918497}.

However, in complex application scenarios, ALAPs are required to provide efficient data communication services, collect and process environmental information, and perform target detection, localization, and imaging.
To meet these demands, integrated sensing and communication (ISAC) has emerged as a spectral-efficient framework, enabling simultaneous optimization of communication capacity and sensing resolution through dual-functional waveform design, dynamic beam management, and intelligent task-oriented resource slicing \cite{9606831}.
ISAC enables multi-dimensional resource integration across wireless communication and radar sensing,
by consolidating both functionalities within the co-designed hardware platforms and shared spectral bands, which not only facilitates the coexistence but also enhances synergy through time-space-frequency resource multiplexing, thereby maximizing resource utilization while optimizing performance \cite{9737357}.
Nowadays, the deep integration of information technology, mobile communications, and artificial intelligence has imposed stringent requirements on information processing capabilities in next-generation wireless networks (NGWNs) \cite{8488502, 9252924}.
To address heterogeneous service requirements, the combination of mobile edge computing (MEC) with ISAC technology, gives rise to the integrated sensing, communication, and computation (ISCC) \cite{10217150, 10709889}.
ISCC enables concurrent delivery of intelligent connectivity, coordinated sensing, and distributed computing services, achieving efficient resource utilization and substantially enhancing system performance while offering a novel perspective for the NGWNs \cite{9397776, 10908649, 10908620}.

The evolution of wireless communications reveals fundamental limitations in technologies like half-duplex and frequency-division duplexing, characterized by inefficient spectrum utilization and suboptimal resource allocation.
Full-duplex (FD) communications effectively overcomes these limitations through advanced self-interference cancellation techniques, enabling bidirectional data transmission within identical frequency bands to improve spectral efficiency \cite{8642523}.
Specifically, FD-enabled systems augment communication and sensing capabilities by efficiently reusing temporal-spectral resource in the ISAC system over ALAP.
From a communication perspective, FD achieves substantial spectral efficiency gains.
From a sensing perspective, persistent sensing can be performed on the FD-enabled ALAP, utilizing all available channel bands, enabling continuous environmental scanning with enhanced sensing resolution.
Nevertheless, despite these advantages, FD operation still faces challenges such as increased hardware complexity and higher energy consumption.

While ISCC technology has demonstrated great potential in enabling multifunctional systems that support simultaneous data transmission, environmental sensing, and edge computing, its integration with ALAP remains underexplored, particularly in complex application scenarios requiring high adaptability and real-time decision-making.
Moreover, although FD communications offer promising gains in spectral efficiency and sensing performance through intelligent resource reuse, their incorporation into ALAP-enabled ISCC systems has not been sufficiently investigated.
These gaps highlight an important research opportunity: how to effectively integrate ISCC and FD technologies into an ALAP framework to jointly optimize communication, sensing, and computing functionalities while addressing the challenges posed by dynamic environments and limited onboard resources.
These consequently motivate our investigation.

This paper studies an FD-ALAP-enabled ISCC system, in which the ALAP has two functions.
On one hand, it can act as a relay between users and the onboard MEC server, receiving task data from users and processing it on the onboard MEC server.
On the other hand, it can perform ground target sensing to address dynamic sensing requirements in complex environments, where the ALAP receives the echo signal at the same time of transmission due to the FD technology.
Under this setup, considering that both ALAP and users incur significant energy consumption for transmission and computation, we jointly optimize resource allocation and coordinated beamforming to minimize the overall system energy consumption, by taking into account of the critical constraints of beamforming, task allocation, computation resource allocation, time allocation, and sensing gain. To address this problem, a novel resource allocation problem is formulated, and an efficient algorithm is proposed.
The main contributions of this work can be summarized as follows:
\begin{itemize}
\item We formulate a new joint optimization problem for the task allocation, the computation resource allocation, the transmitting beamforming, and the receiving beamforming to minimize the total energy consumption of the FD-ALAP-enabled ISCC system. 
Different from the related works, the energy consumption of the entire system is considered, which provides a more comprehensive and accurate model for system performance optimization and evaluation.
\item To address the non-convexity and coupling variables in the optimization problem, we decompose it into four subproblems, with different optimization methods applied based on their characteristics. An alternating optimization (AO)-based algorithm is utilized for obtaining the overall solution.
\item We conduct extensive experiments to validate the effectiveness and robustness of the proposed algorithm. Numerical results demonstrate that the proposed algorithm outperforms multiple benchmark algorithms.
\end{itemize}

\textit{Organizations:} The rest of this paper is organized as follows.
Section II reviews the related works.
Section III introduces the FD-ALAP-enabled ISCC system and formulates the energy consumption minimization problem.
Section IV proposes the AO-based algorithm for obtaining the solution to the problem and analyzes the convergence and complexity of the proposed algorithm.
Section V shows the numerical results of the proposed scheme versus multiple benchmark schemes.
Finally, Section VI concludes the paper.

\textit{Notations:} The notations in this paper are introduced below.
$ {{\mathbb{C}}^{M\times N}} $ denotes the $ M \times N $ complex matrix $ \mathbb{C} $.
$ \mathrm{j} $ represents the imaginary unit, where $ {\mathrm{j}^2} = -1 $.
$\boldsymbol{I}_N$ denotes the $N \times N$ identity matrix.
For a generic matrix $ {\boldsymbol{G}} $, $ {{\boldsymbol{G}}^{\mathsf{H}}} $, $ {{\boldsymbol{G}}^{\mathsf{T}}} $, $ \mathsf{tr}({\boldsymbol{G}}) $, and $\mathsf{rank}({\boldsymbol{G}})$ denote the conjugate transpose, transpose, trace, and rank of $ {\boldsymbol{G}} $, respectively.
$\| \boldsymbol{a} \|$ denotes the norm of $ {\boldsymbol{a}} $.

\section{Related Works}

In this section, we provide a comprehensive review of the existing works.
The following subsections present a detailed discussion from three key perspectives: AAV-enabled communications, ISCC, and FD communications.
These three areas collectively inform the technical background and design considerations of our proposed framework.

\subsection{AAV-enabled Communications}
AAVs, due to their flexibility and deployability, can be rapidly deployed in areas that require temporary enhancement of network coverage or serve as mobile base stations to address emergency situations, such as rescue operations following natural disasters or temporary communication support during large-scale events.
By strategically planning the positioning, flight trajectory, and resource allocation of AAVs, the coverage range of network, data transmission rate, and quality of service can be significantly enhanced \cite{9456851}.
Specifically, the authors in \cite{9712630} and \cite{9604506} studied the AAV-assisted data collection systems for Internet of things (IoTs) devices while optimizing the mobility of the AAV.
The authors in \cite{9672750} focused on the resource allocation and trajectory design for a multiple-input single-output AAV-assisted MEC network, with an emphasis on minimizing the system energy consumption.
The authors in \cite{9273074} considered a multi-AAV-assisted IoTs network and proposed a joint optimization of AAV 3D trajectory and time allocation to maximize the data collection rate under a realistic probabilistic LoS channel model.
The authors in \cite{8937793} focused on an AAV-assisted wireless-powered cooperative MEC system, aiming to minimize the total energy consumption of the AAV by jointly optimizing the task offloading, central processing unit (CPU) control, and trajectory planning.
The authors in \cite{10606316} investigated an AAV-enabled heterogeneous MEC system, aiming to maximize the minimum task computation data volume for active devices, while considering factors such as the AAV trajectory, computation resource allocation, and time allocation.
However, existing AAV-enabled systems mainly considered the singular functionality of information transmission and processing, failing to effectively address complex scenarios requiring environmental sensing capabilities.

\subsection{ISCC}
Recent studies on ISCC have primarily focused on integrating traditional ISAC systems with MEC to simultaneously leverage the advantages of communication, sensing, and computation.
Specifically, the authors in \cite{10757511} studied an ISCC system, aiming to minimize device energy consumption by adjusting beamforming, task allocation, and phase duration allocation.
Furthermore, the authors in \cite{10678874} utilized deep neural networks to optimize network resource utilization and task latency in ISCC.
Driven by the synergistic capabilities of the ISCC, several pioneering studies have explored the combination of ISCC with AAVs \cite{10045764,10233771,10107972,10579910}.
The authors in \cite{10045764} comprehensively characterized the trade-off relationship between the computing capability and beamforming gain by optimizing the AAV's beamforming and flight trajectory.
The authors in \cite{10233771} studied an age of information minimization problem in an AAV-enabled marine ISCC system, and proposed a learning-based scheme for getting the near-optimal strategy while ensuring data freshness.
The authors in \cite{10107972} formulated a joint optimization problem focused on AAV energy consumption and data collection time by optimizing the sensing scheduling, power allocation, and trajectory planning.
The authors in \cite{10579910} proposed a fractional optimization problem to maximize the computing efficiency by optimizing the AAV trajectory, beamforming vector, and computation offloading strategy.
However, under the condition of meeting the demands of various complex application scenarios, how to further enhance spectral efficiency and improve the performance of ISCC systems remains a critical bottleneck in the aforementioned works.

\subsection{FD Communications}
The explosive growth of communication demands has positioned spectral efficiency as a critical factor in enhancing system performance for the NGWNs.
The principal advantage of the FD communications lies in its significant improvement in spectral efficiency, which helps reduce communication latency, increase network capacity, and support user access without requiring additional spectrum resources.
Several previous works have demonstrated the performance gain benefited by the FD communications for the ISAC systems \cite{10158711, 10279303}.
Besides, the authors in \cite{9965407} and \cite{10039247} studied the non-orthogonal transmission in an FD ISAC system, with the alleviation of spectrum congestion and enhancement in system performance.
Furthermore, recent research has explored the integration of FD and ISCC technologies to achieve co-optimized spectral and computational resource utilization \cite{10643175, 10681580}.
Specifically, the authors in \cite{10643175} considered a simultaneously transmitting and reflecting reconfigurable intelligent surface-assisted ISCC framework for the Internet of robotic things, and proposed an optimization problem to maximize the total computation rate of decision robots.
The authors in \cite{10681580} investigated a multi-objective sensing framework that integrates FD and non-orthogonal multiple access in a novel ISCC framework to minimize system energy consumption.
However, FD communications lead to increased system complexity and higher energy consumption due to the simultaneously maintaining of both transmission and receiving functions. Therefore, minimizing system energy consumption represents a promising direction for further exploration.

\begin{figure}[!htbp]
	\centering
	\includegraphics[width=0.75\linewidth]{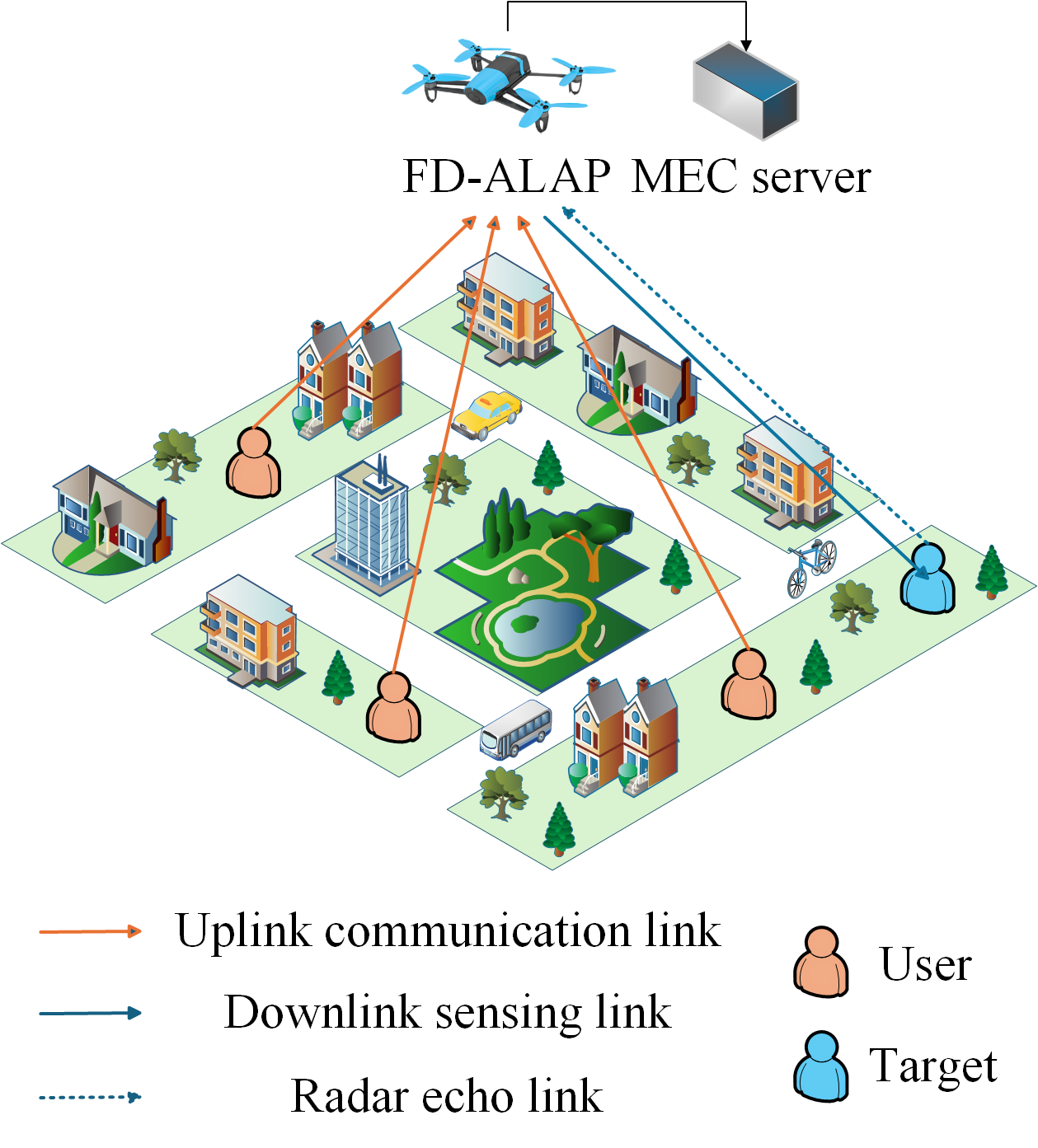}
	\caption{The FD-ALAP-enabled ISCC system.}
	\label{fig:System_model}
\end{figure}

\section{System Model and Problem Formulation}
As illustrated in Fig. \ref{fig:System_model}, we consider an ISCC system, which consists of $M$ single-antenna users, a radar target, and an ALAP capable of FD communications.
The ALAP is equipped with a transmit uniform linear array (ULA), a receive ULA, and an MEC server.
The set of users is denoted as $\mathcal{M}= \{1,2,\cdots,M\}$.
All users face numerous time-intensive tasks requiring significant computational resources but have limited local computing capabilities. Fortunately, they are permitted to offload these tasks to the MEC server via ALAP for efficient task processing services.
Meanwhile, the ALAP conducts radar sensing to detect target-related information, such as classification and features.

We consider a time interval that is segmented into multiple discrete time slots, each with a duration of $\tau$ seconds.
Subsequently, we focus solely on the system scenario within a single time slot.
We assume that, within this time slot, all users and the radar target are stationary on the ground, while the ALAP hovers in the air to ensure optimal performance of communication and target detection.
The distance between the ALAP and the $m$-th user is given by $d_m$, and the distance between the ALAP and the target is denoted as $d_0$.

In the FD-ALAP-enabled ISCC system, the downlink sensing signal, transmitted from the transmit ULA with $N_t$ antennas, is adopted for target detection.
Meanwhile, the radar echo signal and the uplink communication signals are received at the ALAP through the receive ULA with $N_r$ antennas simultaneously.

In the time slot, each user is assigned specific computation tasks that can be divided into two parts and processed concurrently \cite{8387798}.
One portion of the tasks is computed locally by the users. The other portion follows a three-step process: 1) First, it is offloaded to the ALAP; 2) Second, it undergoes processing via the MEC server; 3) Finally, the results are transmitted back to the user.
The data size of the computation results are much smaller compared to the offloaded task data, thus the latency required for transmitting the computing results by the ALAP can be neglected \cite{10443270, 10683327, 10373153}.

\subsection{Communication Model}
The signal transmitted from the $m$-th single-antenna user to the ALAP is given by
\begin{equation}
    x_m = \sqrt{p_m}c_m,
\end{equation}
where $p_m$ and $c_m$ are the transmit power and information symbol of the $m$-th user, respectively. This formulation assumes that the transmit signal is modulated by the square root of the transmit power.

Since the ALAP and the $m$-th user are stationary, and the receive ULA is configured with half-wavelength antenna spacing, the receive array steering vector for the ALAP related to the $m$-th user can be expressed as
\begin{equation}
\label{a_m}
    \boldsymbol{a}_m = \frac{1}{\sqrt{N_r}} \left [ 1, 
    e^{j \pi \sin{\theta_m} }, 
    \cdots, 
    e^{j \pi \left ( N_r - 1 \right ) \sin{\theta_m} } \right ]^\mathsf{T},
\end{equation}
where $\theta_m$ represents the angle between the $m$-th user and the ALAP.

Due to the relatively high altitude of the ALAP, a robust LoS link generally exists between the ALAP and each user, which significantly simplifies the channel modeling process and improves communication reliability \cite{9916163}. The LoS channel vector from the $m$-th user to the ALAP is denoted as
\begin{equation}
    \boldsymbol{h}_m = \sqrt{\beta} d_m^{-1} \boldsymbol{a}_m,
\end{equation}
where $\beta$ is the channel power gain at a reference distance of 1m.
Moreover, considering the fixed position of the radar target and the configuration of the transmit ULA, which employs the same half-wavelength antenna spacing as the receive ULA, the transmit array steering vector and receive array steering vector associated with the target can be expressed as
\begin{equation}
\label{a_t}
    \boldsymbol{a}_t = \frac{1}{\sqrt{N_t}} \left [ 1, 
    e^{j \pi \sin{\theta_0} }, 
    \cdots, 
    e^{j \pi \left ( N_t - 1 \right ) \sin{\theta_0} } \right ]^\mathsf{T},
\end{equation}
\begin{equation}
\label{a_r}
    \boldsymbol{a}_r = \frac{1}{\sqrt{N_r}} \left [ 1, 
    e^{j \pi \sin{\theta_0} }, 
    \cdots, 
    e^{j \pi \left ( N_r - 1 \right )\sin{\theta_0} } \right ]^\mathsf{T},
\end{equation}
where $\theta_0$ is the angle corresponding to the target.

Owing to the FD communications, the ALAP is capable of simultaneously receiving radar echo signals while transmitting sensing signals. In the ISCC system, the ALAP also receives uplink communication signals from users, which contain computation tasks. Consequently, the received signal at the FD-enabled ALAP can be expressed as
\begin{equation}
    \boldsymbol{y} = \sum_{m=1}^M \boldsymbol{h}_m x_m + \beta_0 \boldsymbol{A} \boldsymbol{v} + \boldsymbol{n},
\end{equation}
where $\beta_0 \in \mathbb{C}$ denotes the complex amplitude of the target, primarily determined by the path loss and the radar cross-section, $\boldsymbol{A} \triangleq \boldsymbol{a}_r \boldsymbol{a}_t^\mathsf{H}$, $\boldsymbol{v} \in \mathbb{C}^{N_{t}\times 1}$ represents a dedicated radar signal transmitted by the ALAP to enhance sensing performance, and $\boldsymbol{n} \in \mathbb{C}^{N_r \times 1}$ is additive white Gaussian noise with covariance $\sigma_r^2 \boldsymbol{I}_{N_r}$.
In this study, we assume that the self-interference resulting from FD communications is completely mitigated \cite{10681580}.
By applying a set of receive beamformers ${\{ \boldsymbol{w}_m\} }^M_{m=1} \in \mathbb{C}^{N_r \times 1}$ on the ALAP to recover data signals from uplink users, the receive signal-to-interference-plus-noise ratio for the $m$-th user can be given by
\begin{equation}
\label{SINR}
\begin{split}
    & \gamma_m = \\
    & \frac{p_m \boldsymbol{w}_m^\mathsf{H} \boldsymbol{h}_m \boldsymbol{h}_m^\mathsf{H} \boldsymbol{w}_m}{\boldsymbol{w}_m^\mathsf{H} \left ( \sum_{j=1,j\neq m}^M p_j \boldsymbol{h}_j \boldsymbol{h}_j^\mathsf{H} + \beta_0^2 \boldsymbol{A} \boldsymbol{V} \boldsymbol{A}^\mathsf{H} + \sigma_r^2 \boldsymbol{I}_{N_r} \right ) \boldsymbol{w}_m},
\end{split}
\end{equation}
where $\boldsymbol{V} \triangleq \boldsymbol{v} \boldsymbol{v}^\mathsf{H}$, and ${\| \boldsymbol{w}_m \|}^2 \leq 1$ denotes the receive power constraint of the ALAP.

\subsection{Computing Model}
In this study, it is assumed the $m$-th user has a total task volume of $L_m$ bits that need to be computed within the given time slot.
A portion of these tasks, denoted as $l_m$, is designated for offloading to the ALAP for processing. The task allocation must satisfy the constraint $0 \leq l_m \leq L_m$.
Moreover, the local computation time of the $m$-th user can be expressed as
\begin{equation}
    T_m^{\mathrm{loc}} = \frac{\phi_m \left ( L_m - l_m \right )}{f_m},
\end{equation}
where $\phi_m$ represents the number of CPU cycles required to process one bit of data by the $m$-th user, and $f_m$ denotes the computation resource allocated to the $m$-th user for local computing. The allocation must satisfy the constraint $0 \leq f_m \leq f_{\max}^m$, where $f_{\max}^m$ denotes the maximum computing capability available to the $m$-th user.
Additionally, the local computation time should not exceed the duration of the time slot, i.e., $T_m^{\mathrm{loc}} \leq \tau$.

The communication rate between the $m$-th user and the ALAP can be written as
\begin{equation}
\label{r}
    r_m = B \log_2 \left ( 1 + \gamma_m \right ),
\end{equation}
where $B$ represents the channel bandwidth. In MEC, the transmission time from the $m$-th user to the ALAP and the computation time at the ALAP can be given by
\begin{equation}
    T_m^{\mathrm{tran}} = \frac{l_m}{r_m},
\end{equation}
\begin{equation}
    T_m^{\mathrm{comp}} = \frac{\phi_\mathrm{A} l_m}{f_m^{\mathrm{A}}},
\end{equation}
where $\phi_\mathrm{A}$ represents the number of CPU cycles required to process one bit of data by the ALAP, and $f_m^{\mathrm{A}} \geq 0$ denotes the computation resource allocated to the data packet requested by the $m$-th user at the ALAP.
The maximum available computation resource of the ALAP is denoted as $f_{\max}^\mathrm{A}$, which must satisfy the constraint $\sum_{m=1}^M f_m^{\mathrm{A}} \leq f_{\max}^\mathrm{A}$.
For the data transmitted to the ALAP for processing, the total time, which includes both the transmission time and computation time, cannot exceed the duration of the time slot, i.e., $T_m^{\mathrm{tran}} + T_m^{\mathrm{comp}} \leq \tau$.

\subsection{Sensing Model}
In the FD-ALAP-enabled ISCC system, we evaluate the radar sensing performance towards potential targets within the region of interest by employing the transmit beampattern gain as the primary metric.
The transmit beampattern gain quantifies the distribution of transmit power in the direction of the target, which is specifically designed to meet the requirements of the sensing tasks.
The transmit beampattern gain directed towards the target location is denoted as
\begin{equation}
    \Gamma = \boldsymbol{a}_t^\mathsf{H} \boldsymbol{V} \boldsymbol{a}_t.
\end{equation}

To ensure that the sensing performance meets the predefined threshold, we impose a constraint $\Gamma \geq \left ( d_0 \right )^2 \Gamma_{\min}$, where $\Gamma_{\min}$ represents the minimum required sensing performance threshold.
The sensing gain threshold requirement is positively related to the square of the distance from the ALAP to the sensing target, ensuring that the radar can maintain adequate performance even the target is at a long distance.

\subsection{Energy Consumption Model}
\paragraph{User Computing Energy Consumption}
In the entire time slot, the total local computing energy consumption of all users can be expressed as
\begin{equation}
    E^{\mathrm{loc}} = \sum_{m=1}^M \kappa_m \left ( f_m \right )^3 T_m^{\mathrm{loc}},
\end{equation}
where $\kappa_m$ denotes the energy efficiency factor for the local computing of the $m$-th user. 

\paragraph{User Transmission Energy Consumption}
The total transmission energy consumption of all users in the entire time slot can be determined by summing the transmission energy consumption of each user, i.e.,
\begin{equation}
    E^{\mathrm{tran}} = \sum_{m=1}^M p_m T_m^{\mathrm{tran}}.
\end{equation}

\paragraph{ALAP Computing Energy Consumption}
The total computing energy consumption of the ALAP accounts for the tasks offloaded from all users, which can be written as
\begin{equation}
    E_{\mathrm{A}}^{\mathrm{comp}} = \sum_{m=1}^M \kappa_\mathrm{A} \left ( f_m^\mathrm{A} \right )^3 T_m^{\mathrm{comp}},
\end{equation}
where $\kappa_\mathrm{A}$ denotes the energy efficiency factor for the ALAP.

\paragraph{ALAP Transmission Energy Consumption}
In the overall system, the ALAP is specified to transmit only sensing signals to the target. Therefore, the transmission energy consumption of the ALAP can be defined as
\begin{equation}
    E_{\mathrm{A}}^{\mathrm{tran}} = \tau {\| \boldsymbol{v} \|}^2,
\end{equation}
where the term ${\| \boldsymbol{v} \|}^2$ denotes the power of the transmit signal.

\subsection{Problem Formulation}
We aim to minimize the total energy consumption within the considered FD-ALAP-enabled ISCC system, including the local computing and transmission energy consumption of each user, as well as the computing and transmission energy consumption of the ALAP.
By jointly optimizing the beamforming vector at the ALAP transmitter $\boldsymbol{v}$, the beamforming vectors at the ALAP receiver, $\boldsymbol{W} \triangleq \{ \boldsymbol{w}_m, \forall m \}$, the task allocation of users to the ALAP, $\boldsymbol{L} \triangleq \{ l_m, \forall m \}$, and the computation resource allocation of users and the ALAP, $\boldsymbol{F} \triangleq \{ f_m, f_m^{\mathrm{A}}, \forall m \}$, the overall energy consumption minimization problem can be formulated as
\begin{subequations}
	\begin{align}
\textbf{P1}:\ &\min\limits_{\{\boldsymbol{v}, \boldsymbol{W}, \boldsymbol{L}, \boldsymbol{F}\}} \ E^{\mathrm{loc}} + E^{\mathrm{comp}} + E_{\mathrm{A}}^{\mathrm{comp}} + E_{\mathrm{A}}^{\mathrm{tran}} \nonumber\\
{\mathrm{s.t.}}\  
&{\| \boldsymbol{w}_m \|}^2 \leq 1, \ \forall m, \label{p1a}\\
&0 \leq l_m \leq L_m, \ \forall m, \label{p1b}\\
&0 \leq f_m \leq f_{\max}^m, \ \forall m, \label{p1c}\\
&0 \leq f_m^{\mathrm{A}}, \ \sum_{m=1}^M f_m^{\mathrm{A}} \leq f_{\max}^\mathrm{A}, \ \forall m, \label{p1d}\\
&T_m^{\mathrm{loc}} \leq \tau, \ \forall m, \label{p1e}\\
&T_m^{\mathrm{tran}} + T_m^{\mathrm{comp}} \leq \tau, \ \forall m, \label{p1f}\\
&\Gamma \geq \left ( d_0 \right )^2 \Gamma_{\min}, \label{p1g}
	\end{align}
\end{subequations}
where (\ref{p1a}) represents the beamforming constraint at the receiving ends of ALAP, ensuring that the beamforming vectors are properly designed, 
(\ref{p1b}) represents the task allocation decision constraint of each user, ensuring that the tasks are allocated in a manner that meets the system's operational requirements,
(\ref{p1c}) and (\ref{p1d}) represent the computation resource allocation decision constraints for users and the ALAP, respectively, ensuring that the computation resources are allocated within the available limits,
(\ref{p1e}) and (\ref{p1f}) represent the time constraints, ensuring that all tasks are completed within the specified time intervals,
and (\ref{p1g}) represents the sensing gain constraint, ensuring that the radar sensing performance meets the required threshold.
Problem (\textbf{P1}) is a non-convex optimization problem, which presents significant challenges in identifying the global optimum. The non-convex nature of this problem stems from intricate interactions among multiple constraints and the objective function. These complexities necessitate the application of advanced optimization techniques to achieve effective solutions.

\section{Joint Optimization of Resource Allocation and Coordinated Beamforming}
In this section, we decompose the problem (\textbf{P1}) into four subproblems: task allocation, computation resource allocation, transmitting beamforming, and receiving beamforming. The four subproblems are optimized using an AO manner.

\subsection{Optimization of Task Allocation}
Given $\{\boldsymbol{v}, \boldsymbol{W}, \boldsymbol{F}\}$, the problem of optimizing the task allocation of users to the ALAP $\boldsymbol{L}$ can be formulated as
\begin{subequations}
	\begin{align}
\textbf{P2}:\ &\min\limits_{\{\boldsymbol{L}\}} \ \sum_{m=1}^M \lambda_m l_m + \sum_{m=1}^M \kappa_m \left ( f_m \right )^2 \phi_m L_m + E_{\mathrm{A}}^{\mathrm{tran}} \nonumber\\
{\mathrm{s.t.}}\  
&\text{(\ref{p1b}), (\ref{p1e}) and (\ref{p1f})}, \nonumber
	\end{align}
\end{subequations}
where $\lambda_m = \frac{p_m}{r_m} + \kappa_\mathrm{A} \left ( f_m^\mathrm{A} \right )^2 \phi_\mathrm{A} - \kappa_m \left ( f_m \right )^2 \phi_m$.

To further analyze the problem (\textbf{P2}), it can be rigorously verified that the problem constitutes a linear programming (LP) problem. Given its LP structure, the problem can be efficiently solved using the CVX solver. This approach ensures both accuracy and computational efficiency in finding the optimal solution.

\subsection{Optimization of Computation Resource Allocation}
Given $\{\boldsymbol{v}, \boldsymbol{W}, \boldsymbol{L}\}$, the problem of optimizing the computation resource allocation of users and the ALAP $\boldsymbol{F}$ can be formulated as
\begin{subequations}
	\begin{align}
\textbf{P3}:\ &\min\limits_{\{\boldsymbol{F}\}} \ \sum_{m=1}^M \left [ \mu_m \left ( f_m \right )^2 + \nu_m \left ( f_m^\mathrm{A} \right )^2 \right ] + E^{\mathrm{tran}} + E_{\mathrm{A}}^{\mathrm{tran}} \nonumber\\
{\mathrm{s.t.}}\  
&\text{(\ref{p1c}), (\ref{p1d}), (\ref{p1e}) and (\ref{p1f})}, \nonumber
	\end{align}
\end{subequations}
where $\mu_m = \kappa_m \phi_m \left ( L_m - l_m \right )$ and $\nu_m = \kappa_\mathrm{A} \phi_\mathrm{A} l_m$.

It can be rigorously demonstrated that the functions $\frac{1}{f_m}$ and $\left ( f_m \right )^2$ are convex with respect to $f_m$. Similarly, the functions $\frac{1}{f_m^\mathrm{A}}$ and $\left ( f_m^\mathrm{A} \right )^2$ exhibit convexity with respect to $f_m^\mathrm{A}$. These properties are critical for ensuring the tractability of optimization problems involving these functions, as convexity guarantees that local optima are also global optima.
Hence, the objective function is a convex function, and all constraints are linear. Consequently, the problem (\textbf{P3}) is a convex optimization problem. This characteristic ensures that optimal solutions can be obtained, leveraging the CVX solver.

\subsection{Optimization of Transmitting Beamforming}
Based on our previous definition of $\boldsymbol{V} \triangleq \boldsymbol{v} \boldsymbol{v}^\mathsf{H}$, it follows that $\boldsymbol{V} \succeq 0$ and $\mathsf{rank}(\boldsymbol{V}) \leq 1$.
By substituting the variable with $\boldsymbol{V}$ and given $\{\boldsymbol{W}, \boldsymbol{L}, \boldsymbol{F}\}$, the problem of optimizing the beamforming vector at the ALAP transmitter $\boldsymbol{v}$ can be reformulated as
\begin{subequations}
	\begin{align}
\textbf{P4}:\ &\min\limits_{\{\boldsymbol{V}\}} \ \sum_{m=1}^M \frac{p_m l_m}{r_m} + \tau \mathsf{tr}(\boldsymbol{V}) + E^{\mathrm{loc}} + E_{\mathrm{A}}^{\mathrm{comp}} \nonumber\\
{\mathrm{s.t.}}\ 
&\mathsf{rank}(\boldsymbol{V}) \leq 1, \label{p4a}\\
&\text{(\ref{p1f}) and (\ref{p1g})}. \nonumber
	\end{align}
\end{subequations}

To approximate the problem (\textbf{P4}) and render it more tractable, we introduce auxiliary variables. Specifically, let $R_m = r_m$ represent the auxiliary variable corresponding to $r_m$ and $g_m \leq \beta_0^2 \boldsymbol{w}_m^\mathsf{H} \boldsymbol{A} \boldsymbol{V} \boldsymbol{A}^\mathsf{H} \boldsymbol{w}_m$. 
These substitutions simplify the original problem, facilitating a more efficient solution process.
Furthermore, according to Eq. (\ref{SINR}) and Eq. (\ref{r}), we have
\begin{equation}
\label{p4_1}
    R_m \leq  B \log_2 \left ( 1 + \frac{b_m}{g_m + e_m} \right ),
\end{equation}
where $b_m$ and $e_m$ can be given by
\begin{equation}
    b_m = p_m \boldsymbol{w}_m^\mathsf{H} \boldsymbol{h}_m \boldsymbol{h}_m^\mathsf{H} \boldsymbol{w}_m,
\end{equation}
\begin{equation}
    e_m = \boldsymbol{w}_m^\mathsf{H} \left ( \sum_{j=1,j\neq m}^M p_j \boldsymbol{h}_j \boldsymbol{h}_j^\mathsf{H} + \sigma_r^2 \boldsymbol{I}_{N_r} \right ) \boldsymbol{w}_m.
\end{equation}

It can be shown through standard convexity analysis that the function $B \log_2 \left ( 1 + \frac{b_m}{g_m + e_m} \right )$ is convex with respect to $g_m$.
By applying the successive convex approximation (SCA) method, we approximate the right-hand-side (RHS) of (\ref{p4_1}) in a computationally efficient manner.
Specifically, using the first-order Taylor expansion at the given point $g^{(i)}_m$ during the $i$-th iteration of the approximation process, the lower-bound for the RHS can be derived as
\begin{equation}
\begin{split}
    & B \log_2 \left ( 1 + \frac{b_m}{g_m + e_m} \right ) \geq \\ 
    & B \biggl( \log_2 \left (g_m + e_m + b_m\right )  - \log_2 \left (g^{(i)}_m + e_m\right ) \\
    & - \frac{\log_2(e)}{g^{(i)}_m + e_m} \left (g_m - g^{(i)}_m \right ) \biggr) \triangleq r^\mathrm{lb}_m.
\end{split}
\end{equation}

This approach leverages the iterative nature of the SCA method to progressively refine the approximation, ensuring convergence towards the desired solution.
Therefore, the problem (\textbf{P4}) can be approximated by the following optimization problem:
\begin{subequations}
	\begin{align}
\textbf{P4A}:\ &\min\limits_{\{\boldsymbol{V}, \boldsymbol{R}, \boldsymbol{G} \}} \ \sum_{m=1}^M \frac{p_m l_m}{R_m} + \tau \mathsf{tr}(\boldsymbol{V}) + E^{\mathrm{loc}} + E_{\mathrm{A}}^{\mathrm{comp}} \nonumber\\
{\mathrm{s.t.}}\ 
&\frac{l_m}{R_m} + T_m^{\mathrm{comp}} \leq \tau, \ \forall m, \label{p4Aa}\\
&R_m \leq r^\mathrm{lb}_m, \ \forall m, \label{p4Ab}\\
&g_m \leq \beta_0^2 \boldsymbol{w}_m^\mathsf{H} \boldsymbol{A} \boldsymbol{V} \boldsymbol{A}^\mathsf{H} \boldsymbol{w}_m, \ \forall m, \label{p4Ac}\\
&\text{(\ref{p1g}) and (\ref{p4a})}, \nonumber
	\end{align}
\end{subequations}
where $\boldsymbol{R} \triangleq \{ R_m, \forall m \}$ and $\boldsymbol{G} \triangleq \{ g_m, \forall m \}$.
\begin{remark}
In the problem (\textbf{P4A}), the objective function is convex.
However, the constraint \eqref{p4a} is still non-convex. To address this, we relax the constraint \eqref{p4a} to convert it into a standard convex optimization problem.
Although the relaxed solution may not guarantee rank-one matrices, the rank-one solution can be approximated by Gaussian randomization \cite{9916163, 10654366, 10497119}.
Fortunately, the following proposition shows that an optimal rank-one solution $\boldsymbol{V}^{*}$ to this problem always exists. Hence, Gaussian randomization is not required.
\end{remark}

\begin{proposition}
\label{prop_rank}
There always exists a global optimal solution to the problem (\textbf{P4A}), denoted as $\boldsymbol{V}^{*}$, such that
\begin{equation}
    \mathsf{rank}(\boldsymbol{V}^{*}) = 1.
\end{equation}
\begin{proof}
Please refer to Appendix \ref{app_rank}.
\end{proof}
\end{proposition}

Meanwhile, all other constraints are linear, which, in combination with the convex objective function, ensures that the problem (\textbf{P4A}) is a convex optimization problem.
Consequently, it can be effectively solved using the CVX solver.

\subsection{Optimization of Receiving Beamforming}
According to the conventional approach in receiving beamforming, we first define $\boldsymbol{W}_m = \boldsymbol{w}_m \boldsymbol{w}_m^\mathsf{H}$. Consequently, it follows that $\boldsymbol{W}_m \succeq 0$ and $\mathsf{rank}(\boldsymbol{W}_m) \leq 1$.
By substituting the variables with $\boldsymbol{{W}'} \triangleq \{ \boldsymbol{W}_m \succeq 0, \forall m \}$, and given $\{\boldsymbol{v}, \boldsymbol{L}, \boldsymbol{F}\}$, the problem of optimizing the beamforming vectors at the ALAP receiver $\boldsymbol{W}$ can be reformulated as
\begin{subequations}
	\begin{align}
\textbf{P5}:\ &\min\limits_{\{\boldsymbol{{W}'}\}} \ \sum_{m=1}^M \frac{p_m l_m}{r_m} + E^{\mathrm{loc}} + E_{\mathrm{A}}^{\mathrm{comp}} + E_{\mathrm{A}}^{\mathrm{tran}} \nonumber\\
{\mathrm{s.t.}}\  
&\mathsf{tr}(\boldsymbol{W}_m) \leq 1, \ \forall m, \label{p5a}\\
&\mathsf{rank}(\boldsymbol{W}_m) \leq 1, \ \forall m, \label{p5b}\\
&\text{(\ref{p1f})}. \nonumber
	\end{align}
\end{subequations}

To approximate the problem (\textbf{P5}) and obtain its solution, we introduce an auxiliary variable $S_m \leq r_m$. This auxiliary variable is introduced to simplify the original objective function and enhance computational tractability.
Consequently, based on Eq. (\ref{SINR}) and Eq. (\ref{r}), we have
\begin{equation}
\label{p5_1}
    S_m \leq  B \log_2 \left ( 1 + \frac{\mathsf{tr}(\boldsymbol{\Omega}_m \boldsymbol{W}_m)}{\mathsf{tr}(\boldsymbol{\Lambda}_m \boldsymbol{W}_m)} \right ),
\end{equation}
where $\boldsymbol{\Omega}_m = p_m \boldsymbol{h}_m \boldsymbol{h}_m^\mathsf{H}$ and $\boldsymbol{\Lambda}_m = \sum_{j=1,j\neq m}^M p_j \boldsymbol{h}_j \boldsymbol{h}_j^\mathsf{H} + \beta_0^2 \boldsymbol{A} \boldsymbol{V} \boldsymbol{A}^\mathsf{H} + \sigma_r^2 \boldsymbol{I}_{N_r}$.

It can be verified that the function $B \log_2 \left ( 1 + \frac{\mathsf{tr}(\boldsymbol{\Omega}_m \boldsymbol{W}_m)}{\mathsf{tr}(\boldsymbol{\Lambda}_m \boldsymbol{W}_m)} \right )$ is convex with respect to $\boldsymbol{W}_m$. This result follows from the properties of the trace operator and the logarithmic function, combined with the structure of the given expression.
By applying the SCA method, we approximate the RHS of  (\ref{p5_1}) using a first-order Taylor expansion. Specifically, at the given point $\boldsymbol{W}^{(i)}_m$ during the $i$-th iteration of the approximation process, the lower-bound for the RHS can be derived as
\begin{equation}
\begin{split}
    & B \log_2 \left ( 1 + \frac{\mathsf{tr}(\boldsymbol{\Omega}_m \boldsymbol{W}_m)}{\mathsf{tr}(\boldsymbol{\Lambda}_m \boldsymbol{W}_m)} \right ) \geq \\ 
    & B \biggl( \log_2 \Bigl(\mathsf{tr}\bigl( (\boldsymbol{\Omega}_m + \boldsymbol{\Lambda}_m)\boldsymbol{W}_m \bigr) \Bigr) - \log_2 \left (\mathsf{tr}(\boldsymbol{\Lambda}_m \boldsymbol{W}^{(i)}_m) \right ) \\
    & - \mathsf{tr} \left ( \boldsymbol{\Delta}^{(i)}_m (\boldsymbol{W}_m - \boldsymbol{W}^{(i)}_m) \right ) \biggr) \triangleq \overline{r}^\mathrm{lb}_m,
\end{split}
\end{equation}
where $\boldsymbol{\Delta}^{(i)}_m$ can be given by
\begin{equation}
    \boldsymbol{\Delta}^{(i)}_m = \frac{\log_2(e) \boldsymbol{\Lambda}^\mathsf{H}_m}{\mathsf{tr}(\boldsymbol{\Lambda}_m \boldsymbol{W}^{(i)}_m)}.
\end{equation}

Therefore, by leveraging the approximation method described earlier, the problem (\textbf{P5}) can be reformulated and approximated as the following optimization problem:
\begin{subequations}
	\begin{align}
\textbf{P5A}:\ &\min\limits_{\{\boldsymbol{{W}'}, \boldsymbol{S}\}} \ \sum_{m=1}^M \frac{p_m l_m}{S_m} + E^{\mathrm{loc}} + E_{\mathrm{A}}^{\mathrm{comp}} + E_{\mathrm{A}}^{\mathrm{tran}} \nonumber\\
{\mathrm{s.t.}}\  
&\frac{l_m}{S_m} + T_m^{\mathrm{comp}} \leq \tau, \ \forall m, \label{p5Aa}\\
&S_m \leq \overline{r}^\mathrm{lb}_m, \ \forall m, \label{p5Ab}\\
&\text{(\ref{p5a}) and (\ref{p5b})}, \nonumber
	\end{align}
\end{subequations}
where $\boldsymbol{S} \triangleq \{ S_m, \forall m \}$.

\begin{remark}
To solve the optimization problem (\textbf{P5A}), it is observed that while the objective function exhibits convexity, the constraint \eqref{p5b} remains non-convex. To address this issue, we opt to relax the constraint \eqref{p5b}, thereby transforming the original problem into a standard convex problem.
Meanwhile, the rank-one solution can be effectively approximated through Gaussian randomization.
Then, the solution to the relaxed problem is also the solution to the problem (\textbf{P5A}).
\end{remark}
Then, the objective function is convex, while all other constraints are linear.
This ensures that the problem (\textbf{P5A}) is a convex optimization problem. Consequently, it can be effectively solved using the CVX ToolBox.

\subsection{AO-based Algorithm}
\begin{figure}[!htbp]
	\centering
	\includegraphics[width=1\linewidth]{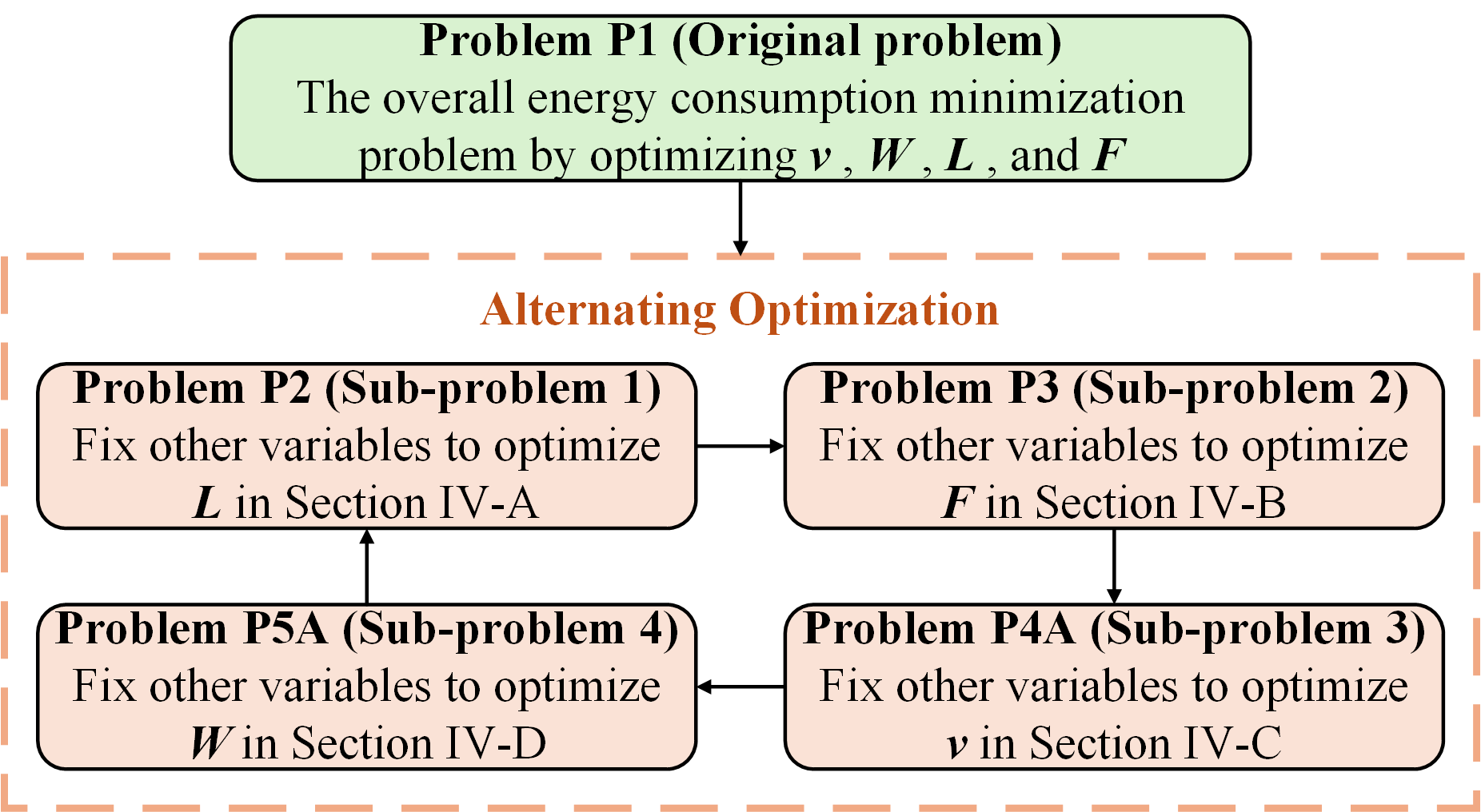}
	\caption{The sequential steps of Algorithm 1.}
	\label{fig:Flow}
\end{figure}

The AO-based algorithm for solving the problem (\textbf{P1}) is detailed in Algorithm \ref{Alg1}.
In lines $1-2$, the value of the objective function and the optimization variables are initialized to random values that satisfy all constraints, and the correlation coefficient for the iteration is also initialized.
Subsequently, we optimize task allocation, computation resource allocation, transmit beamforming, and receive beamforming in an alternating fashion in lines $3-9$.
Then, the process continues until the value of the objective function converges or the maximum iteration number is reached in line $10$.
The sequential steps of Algorithm \ref{Alg1} is shown in Fig.~\ref{fig:Flow}.

\begin{algorithm} \small
\caption{AO-based Algorithm for Solving ($\textbf{P1}$)}
\label{Alg1}
\begin{algorithmic}[1]
    \REQUIRE An initial feasible point $\{\boldsymbol{v}^0, \boldsymbol{W}^0, \boldsymbol{L}^0, \boldsymbol{F}^0\}$;
    \STATE \textbf{Initialize:} Iteration number $i=0$, precision threshold $\epsilon$ and the number of maximum iterations $i_{\max}$;
    \STATE Compute the initial objective function value, i.e. $\Phi(\boldsymbol{v}^0, \boldsymbol{W}^0, \boldsymbol{L}^0, \boldsymbol{F}^0)$;
    \REPEAT
    \STATE Solve the problem (\textbf{P2}) to get the task allocation of users to the ALAP $\boldsymbol{L}^{i+1}$ for given $\{\boldsymbol{v}^i, \boldsymbol{W}^i, \boldsymbol{F}^i\}$;
    \STATE Solve the problem (\textbf{P3}) to get the computation resource allocation of users and the ALAP $\boldsymbol{F}^{i+1}$ for given $\{\boldsymbol{v}^i, \boldsymbol{W}^i, \boldsymbol{L}^{i+1}\}$;
    \STATE Solve the problem (\textbf{P4A}) to get the beamforming vector at the ALAP transmitter $\boldsymbol{v}^{i+1}$ for given $\{\boldsymbol{W}^{i}, \boldsymbol{L}^{i+1}, \boldsymbol{F}^{i+1}\}$;    
    \STATE Solve the problem (\textbf{P5A}) to get the beamforming vectors at the ALAP receiver $\boldsymbol{W}^{i+1}$ for given $\{\boldsymbol{v}^{i+1}, \boldsymbol{L}^{i+1}, \boldsymbol{F}^{i+1}\}$;
    \STATE Update the objective function value according to above variables, i.e. $\Phi(\boldsymbol{v}^{i+1}, \boldsymbol{W}^{i+1}, \boldsymbol{L}^{i+1}, \boldsymbol{F}^{i+1})$;
    \STATE Update $i = i+1$;
    \UNTIL The objective function between two adjacent iterations is smaller than precision threshold $\epsilon$ or $i > i_{\max}$;
    \ENSURE $\Phi(\boldsymbol{v}^*, \boldsymbol{W}^*, \boldsymbol{L}^*, \boldsymbol{F}^*)$, $\boldsymbol{v}^*$, $\boldsymbol{W}^*$, $\boldsymbol{L}^*$ and $\boldsymbol{F}^*$.
\end{algorithmic}
\end{algorithm}

\paragraph{Convergence Analysis}
The following proposition analyzes the convergence of Algorithm 1.

\begin{proposition}
\label{prop_conv}
The objective function of the problem (\textbf{P1}) is non-increasing with the increase in the number of iterations. Therefore, Algorithm 1 is guaranteed to converge.
\begin{proof}
Please refer to Appendix \ref{app_conv}.
\end{proof}
\end{proposition}

\paragraph{Complexity Analysis}
The aforementioned four sub-problems are solved using the interior-point method via the CVX ToolBox in MATLAB. The complexity of the four sub-problems is $\mathcal{O}(\log(\epsilon^{-1}) (M)^{3.5})$, $\mathcal{O}(\log(\epsilon^{-1}) (2M)^{3.5})$, $\mathcal{O}(\log(\epsilon^{-1}) (N_t^2)^{3.5})$, and $\mathcal{O}(\log(\epsilon^{-1}) (M N_r^2)^{3.5})$, respectively, where $\epsilon$ represents the stopping tolerance. Consequently, the computational complexity of Algorithm 1 in the worst case is $\mathcal{O}(i\log(\epsilon^{-1}) ((M)^{3.5} + (2M)^{3.5} + (N_t^2)^{3.5} + (M N_r^2)^{3.5}))$, where $i$ denotes the iteration number.

\section{Numerical Results}
\begin{figure}[!htbp]
	\centering
	\includegraphics[width=0.8\linewidth]{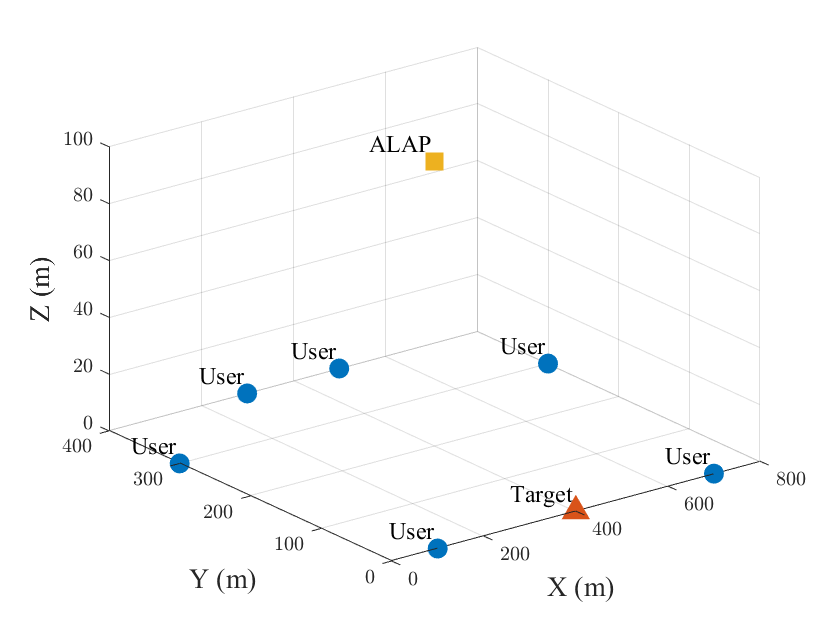}
	\caption{Simulation setup for the ISCC system.}
	\label{fig:Simulation_setup}
\end{figure}

In this section, we present numerical results to validate the effectiveness of the proposed scheme.
The simulation considers a region with dimensions of $400\mathrm{m} \times 800\mathrm{m} \times 100\mathrm{m}$, encompassing $M = 4$ users and a target within this area.
Additionally, the ALAP is deployed at a low altitude of 100 meters for communication, computation, and sensing.
The detailed experimental setup is illustrated in Fig.~\ref{fig:Simulation_setup}.
Other parameters of the experiment are set as follows.
The time slot is $\tau = 2\mathrm{s}$.
The transmit ULA has $N_t = 6$ antennas and the receive ULA has $N_r = 6$ antennas.
The constant channel gain is $\beta = -60\mathrm{dB}$.
The noise power is $\sigma_r^2 = -110\mathrm{dBm}$.
The complex amplitude of the target is $\beta_0^2 = -110\mathrm{dBm}$.
The numbers of CPU cycles required for processing one bit data by users and the ALAP are $\phi_m = 100$ and $\phi_\mathrm{A} = 50$.
The available computation resources of users and the ALAP are $f_{\max}^m = 4 \times 10^6$ CPU cycles/s and $f_{\max}^\mathrm{A} = 8 \times 10^7$ CPU cycles/s.
The channel bandwidth is $B = 500\mathrm{kHz}$.
The energy efficiency factor of the users and the ALAP are $\kappa_m = 10^{-20}$ and $\kappa_\mathrm{A} = 10^{-20}$.

\begin{figure}[!htbp]
	\centering
	\includegraphics[width=0.8\linewidth]{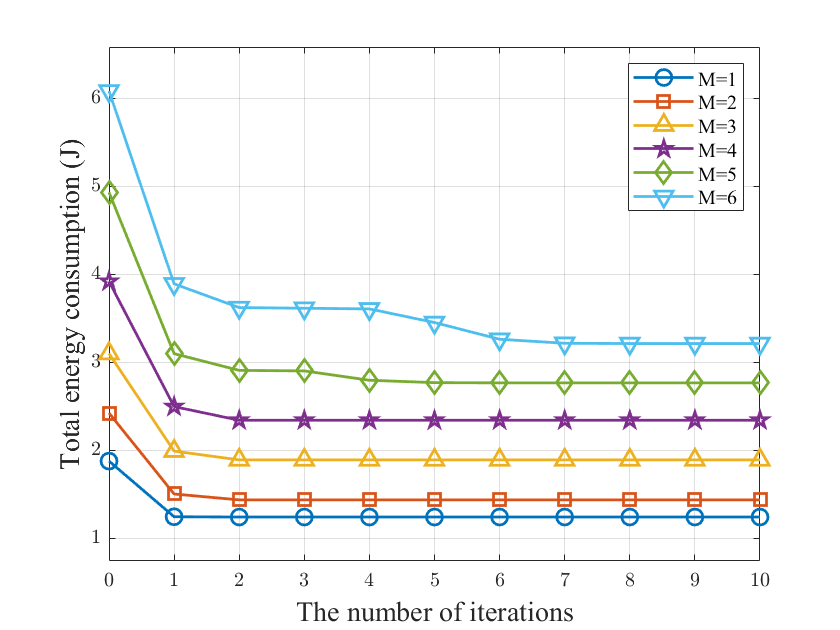}
	\caption{Total energy consumption versus the number of iterations with different numbers of users.}
	\label{fig:Convergence}
\end{figure}
Fig.~\ref{fig:Convergence} depicts the relationship between the total energy consumption and the number of iterations under varying numbers of users.
The results demonstrate that our proposed scheme can converge with fewer iterations, exhibiting a favorable convergence rate.
Specifically, as the number of users increases, the overall task volume grows, and the optimization of beamforming grows more complex, resulting in a higher number of iterations required to achieve convergence.
Overall, the scenarios are able to achieve convergence after eight iterations.

\begin{figure}[!htbp]
	\centering
	\includegraphics[width=0.8\linewidth]{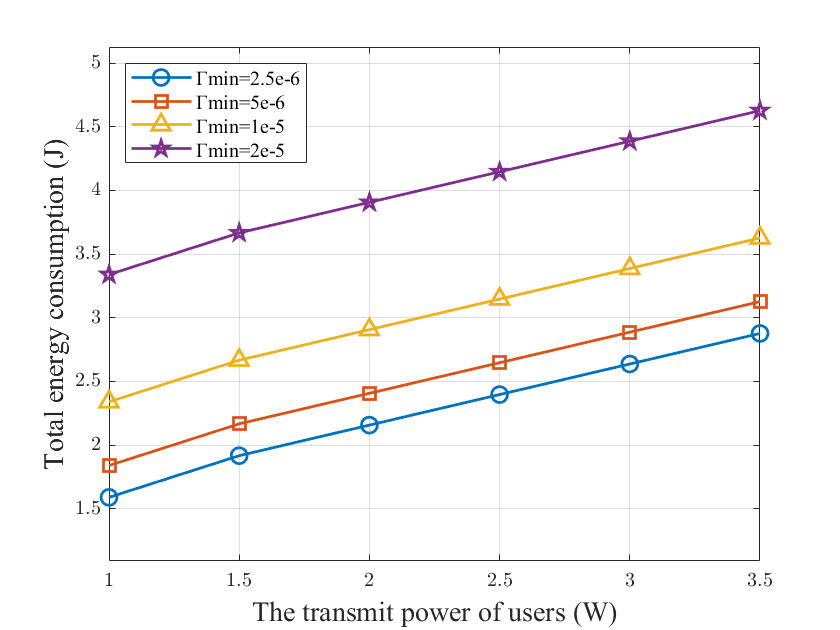}
	\caption{Total energy consumption versus the transmit power of users with different predetermined thresholds.}
	\label{fig:p_Gamma}
\end{figure}
The total energy consumption versus the transmit power of users with different predetermined thresholds is illustrated in Fig.~\ref{fig:p_Gamma}.
As we can observe, the total energy consumption increases as the transmit power of users increases, with other conditions fixed. This is primarily due to the increase in transmission energy consumption of the users.
When the transmit power of users is fixed, the total energy consumption increases with the increase of the minimum required sensing performance threshold. This is because the transmit beamforming vector of ALAP will increase to meet the radar sensing performance, thus increasing the energy consumption.

\begin{figure}[!htbp]
	\centering
	\includegraphics[width=0.8\linewidth]{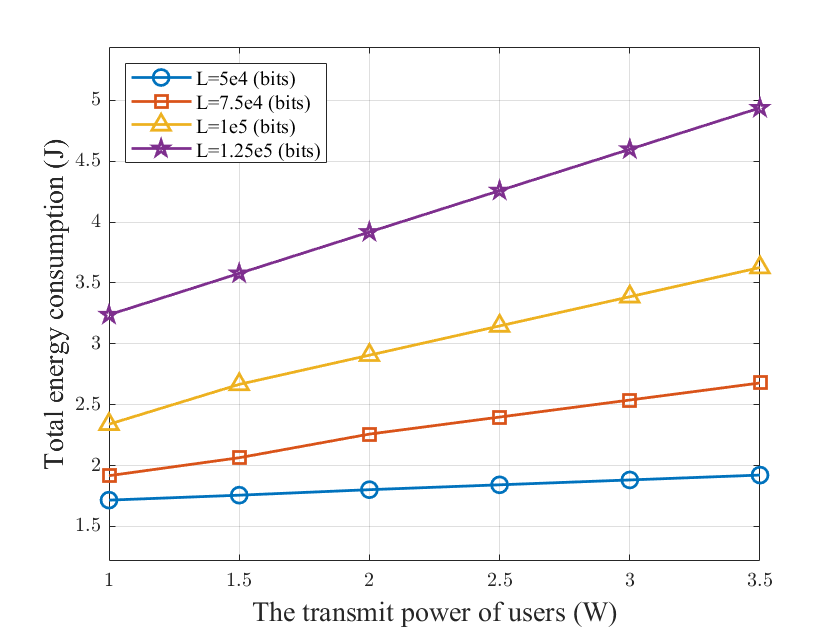}
	\caption{Total energy consumption versus the transmit power of users with different total task volume of users.}
	\label{fig:p_L}
\end{figure}
In Fig.~\ref{fig:p_L}, the total energy consumption versus the transmit power of users with different task volume is presented.
We can observe that, with a constant transmit power, the total energy consumption increases as the number of tasks grows. This is because more energy is required to transmit and compute the additional tasks.
When the number of tasks increases, the total energy consumption increases more obviously with the transmit power. The reason is that the users offloads more tasks to the ALAP with the increased number of tasks, resulting in increased energy consumption.
When the number of tasks is relatively small, the total energy consumption is less influenced by the communication capability and the transmit power.

\begin{figure}[!htbp]
	\centering
	\includegraphics[width=0.8\linewidth]{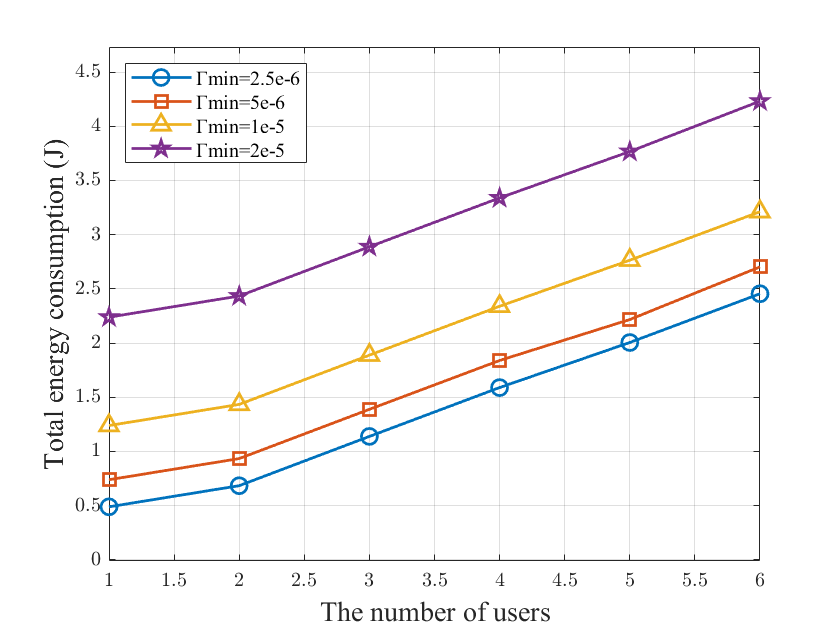}
	\caption{Total energy consumption versus the number of users with different predetermined thresholds.}
	\label{fig:M_Gamma}
\end{figure}
In Fig.~\ref{fig:M_Gamma}, the relationship between the total energy consumption and the number of users is presented, considering different predetermined thresholds.
The results demonstrate that, under different sensing thresholds, the total energy consumption increases as the number of users grows, with a generally consistent increasing rate. This is because a higher number of users leads to a greater overall task volume in the system. Since the task volume assigned to each user remains constant, the increase in energy consumption tends to be linear.
Meanwhile, for a fixed number of users, a higher predetermined threshold results in greater total energy consumption, primarily due to the influence of sensing capability.

The following benchmark schemes are considered for comparisons:
1) \textit{Fixed-task allocation} (FT). The task allocation of the scheme is fixed.
2) \textit{Random-task allocation} (RT). The scheme assigns tasks randomly.
3) \textit{Fixed-computation resource allocation} (FC). The scheme has fixed computation resource allocation.
4) \textit{Random-computation resource allocation} (RC). The computation resource allocation of the scheme is random within a certain range.
5) \textit{Random-beamforming} (RB). In this scheme, the beamformings are randomly generated while the other variables are optimized.

\begin{figure}[!htbp]
	\centering
	\includegraphics[width=0.8\linewidth]{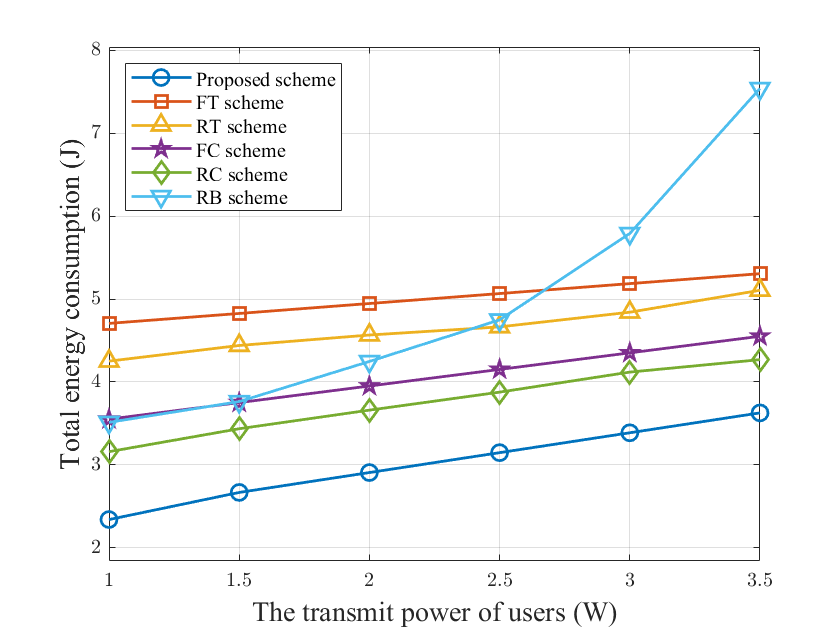}
	\caption{Total energy consumption versus the transmit power of users with different schemes.}
	\label{fig:p_scheme}
\end{figure}
The total energy consumption versus the transmit power of users with different schemes is presented in Fig.~\ref{fig:p_scheme}.
We can observe that the proposed scheme always achieves the lowest energy consumption compared to all benchmarks due to the efficient optimization of task allocation and computation resource allocation as well as beamforming. This verifies that all optimization variables in the proposed scheme have a beneficial effect on the results.
Specifically, because of the randomness of the beamforming vectors, the RB scheme greatly affects the communication rate between the user and the ALAP, and thus produces poor results when the transmit power of users is large.

\begin{figure}[!htbp]
	\centering
	\includegraphics[width=0.8\linewidth]{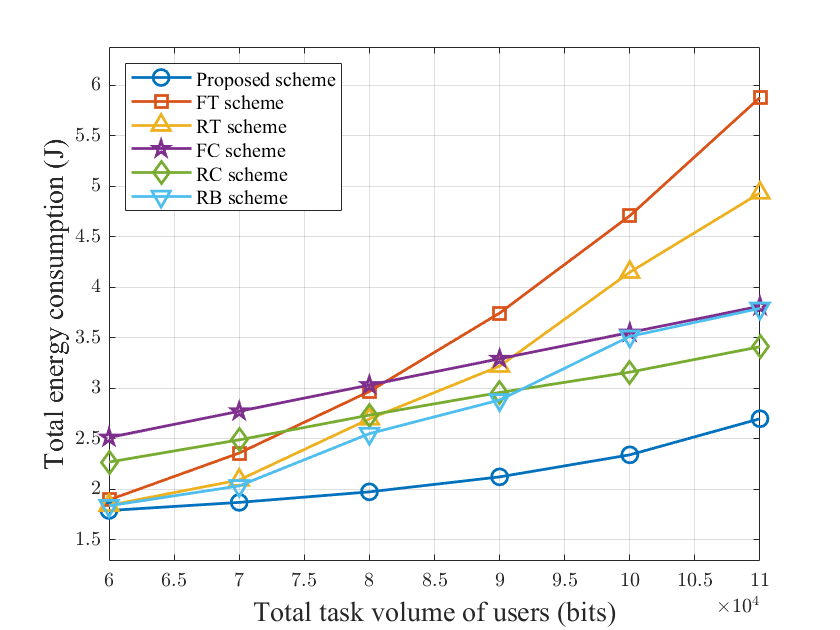}
	\caption{Total energy consumption versus total task volume of users with different schemes.}
	\label{fig:L_scheme}
\end{figure}
Fig.~\ref{fig:L_scheme} illustrates the total energy consumption versus the total task volume of users with different schemes.
This figure shows that the proposed scheme can still achieve the minimum energy consumption for different total task volume.
As the task volume increases from a small level, the superior performance of the proposed scheme is increasingly evident compared to other schemes.
In particular, compared with the FT and RT schemes, the impact of task allocation optimization is less significant when the task volume is small, resulting in comparable total energy consumption. However, as the number of tasks increases, the optimization of task allocation grows more critical, demonstrating the superior performance of the proposed scheme.

\begin{figure}[!htbp]
	\centering
	\includegraphics[width=0.8\linewidth]{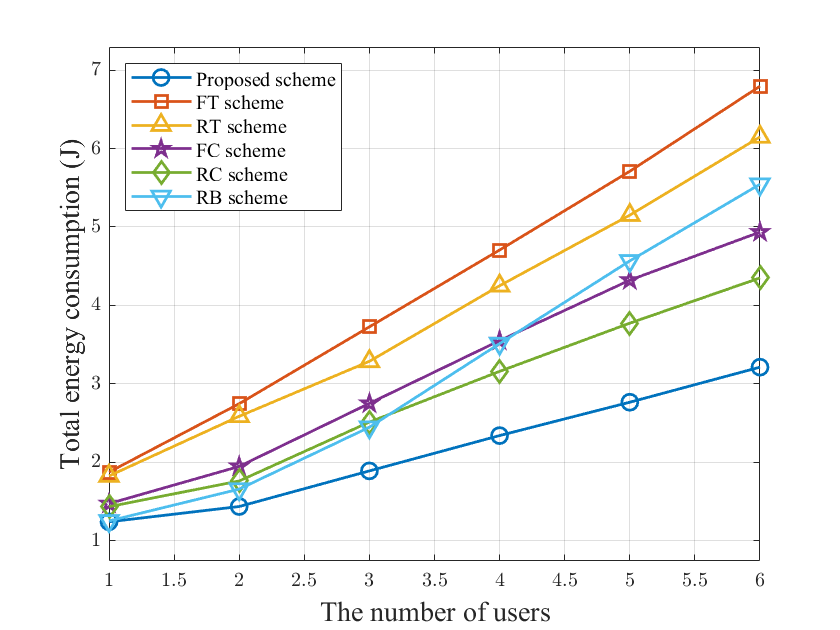}
	\caption{Total energy consumption versus the number of users with different schemes.}
	\label{fig:M_scheme}
\end{figure}
Fig.~\ref{fig:M_scheme} presents the comparison of the total energy consumption across varying numbers of users for different schemes.
As we can observe, under all user quantity scenarios, our proposed scheme still achieves the lowest total energy consumption.
Compared with all benchmark schemes, the optimization performance of our proposed scheme improves as the number of users increases. This is because the increase in the number of users leads to a higher overall task volume, requiring the system to focus more on its communication and computation capabilities. Consequently, the criticality of optimizing task allocation and computation resource allocation escalates. Additionally, the growing number of users increases the complexity of the system, making the optimization of beamforming more challenging.

\section{Conclusion}
This paper investigates an FD-ALAP-enabled ISCC system, where the total energy consumption is minimized by jointly optimizing the beamforming vector at the ALAP transmitter, the beamforming vectors at the ALAP receiver, the task allocation of users to the ALAP, and the computation resource allocation of users and the ALAP.
The numerical results of the proposed scheme confirm the significant performance improvement in terms of energy consumption, as compared to five benchmark schemes.

This paper focuses on the fixed positioning of the ALAP. Future work could further investigate maneuver control of the ALAP to fully leverage the mobility of the AAV.
Additionally, future work could also explore scenarios involving multiple sensing targets or investigate more complex MEC environments that incorporate downlink users.

\begin{appendices}
\section{Proof of Proposition \ref{prop_rank}}
\label{app_rank}
\begin{proof}
Given the problem (\textbf{P4A}) constitutes a convex semidefinite program and satisfies Slater's condition, it follows that the duality gap is zero. Consequently, the Karush-Kuhn-Tucker (KKT) conditions are both necessary and sufficient criteria for ensuring optimality.
Let $\{ \upsilon^{\mathrm{L1}}_m \geq 0 \}^{M}_{m=1}$, $\{ \upsilon^{\mathrm{L2}}_m \geq 0 \}^{M}_{m=1}$, $\{ \upsilon^{\mathrm{L3}}_m \geq 0 \}^{M}_{m=1}$, $\upsilon^{\mathrm{L4}} \geq 0$, and a positive semidefinite matrix $\boldsymbol{Z} \succeq 0$ denote the Lagrange multipliers associated with the constraints (\ref{p4Aa}), (\ref{p4Ab}), (\ref{p4Ac}), (\ref{p1g}) and the semidefinition constraint $\boldsymbol{V} \succeq 0$, respectively. Therefore, the partial Lagrangian function for the problem (\textbf{P4A}) can be formulated as
\begin{equation}
\begin{split}
    & \mathcal{L}(\boldsymbol{V}, \boldsymbol{R}, \boldsymbol{G}, \{ \upsilon^{\mathrm{L1}}_m, \upsilon^{\mathrm{L2}}_m, \upsilon^{\mathrm{L3}}_m\}^{M}_{m=1}, \upsilon^{\mathrm{L4}}, \boldsymbol{Z}) \\
    & = \sum_{m=1}^M \frac{p_m l_m}{R_m} + \tau \mathsf{tr}(\boldsymbol{V}) + E^\mathrm{sum4} \\
    & + \sum_{m=1}^M \upsilon^{\mathrm{L1}}_m \left ( \frac{l_m}{R_m} + T_m^c - \tau \right ) + \sum_{m=1}^M \upsilon^{\mathrm{L2}}_m \left ( R_m - r^\mathrm{lb}_m \right ) \\
    & + \sum_{m=1}^M \upsilon^{\mathrm{L3}}_m \left ( \beta_0^2 \boldsymbol{w}_m^\mathsf{H} \boldsymbol{A} \boldsymbol{V} \boldsymbol{A}^\mathsf{H} \boldsymbol{w}_m - g_m \right ) \\
    & + \upsilon^{\mathrm{L4}} \left ( \boldsymbol{a}_t^\mathsf{H} \boldsymbol{V} \boldsymbol{a}_t - \left ( d_0 \right )^2 \Gamma_{\min} \right ) - \mathsf{tr}(\boldsymbol{Z} \boldsymbol{V}).
\end{split}
\end{equation}
With the Lagrangian function, we further derive the dual function of the problem (\textbf{P4A}) by
\begin{equation}
\begin{split}
    & f(\{ \upsilon^{\mathrm{L1}}_m, \upsilon^{\mathrm{L2}}_m, \upsilon^{\mathrm{L3}}_m\}^{M}_{m=1}, \upsilon^{\mathrm{L4}}, \boldsymbol{Z}) \\
    & = \inf\limits_{\{\boldsymbol{V}, \boldsymbol{R}, \boldsymbol{G} \}} \mathcal{L}(\boldsymbol{V}, \boldsymbol{R}, \boldsymbol{G}, \{ \upsilon^{\mathrm{L1}}_m, \upsilon^{\mathrm{L2}}_m, \upsilon^{\mathrm{L3}}_m\}^{M}_{m=1}, \upsilon^{\mathrm{L4}}, \boldsymbol{Z}) \\
    & = \inf\limits_{\{\boldsymbol{V}, \boldsymbol{R}, \boldsymbol{G} \}} \mathsf{tr}(\boldsymbol{D} \boldsymbol{V}) + h(\boldsymbol{R}, \boldsymbol{G}),
\end{split}
\end{equation}
where $\boldsymbol{D} = \tau \boldsymbol{I}_{N_t} + \sum_{m=1}^M \upsilon^{\mathrm{L3}}_m \boldsymbol{A}^\mathsf{H} \boldsymbol{w}_m \boldsymbol{w}_m^\mathsf{H} \boldsymbol{A} + \upsilon^{\mathrm{L4}} \boldsymbol{a}_t \boldsymbol{a}_t^\mathsf{H} - \boldsymbol{Z}$ and $h(\boldsymbol{R}, \boldsymbol{G})$ represents the remaining terms related to $\{\boldsymbol{R}, \boldsymbol{G} \}$.
$\boldsymbol{D} = 0$ ensures the existence of a bounded dual optimal value, which means that
\begin{equation}
\label{Z}
    \boldsymbol{Z} = \tau \boldsymbol{I}_{N_t} + \sum_{m=1}^M \upsilon^{\mathrm{L3}}_m \boldsymbol{A}^\mathsf{H} \boldsymbol{w}_m \boldsymbol{w}_m^\mathsf{H} \boldsymbol{A} + \upsilon^{\mathrm{L4}} \boldsymbol{a}_t \boldsymbol{a}_t^\mathsf{H}.
\end{equation}
Furthermore, due to the non-negativeness of the Lagrange multipliers, based on Eq. (\ref{Z}), we can conclude that $\mathsf{rank}(\boldsymbol{Z}) \geq N_t - 1$.
Conversely, we present the relevant partial KKT conditions for the problem (\textbf{P4A}) as follows
\begin{equation}
\label{ZWt}
    \boldsymbol{Z}^{*} \boldsymbol{V}^{*} = 0.
\end{equation}
\begin{equation}
\label{Wt}
    \boldsymbol{a}_t^\mathsf{H} \boldsymbol{V}^{*} \boldsymbol{a}_t \geq \left ( d_0 \right )^2 \Gamma_{\min}.
\end{equation}
By jointly considering $\mathsf{rank}(\boldsymbol{Z}) \geq N_t - 1$ and Eq. (\ref{ZWt}), we can conclude that $\mathsf{rank}(\boldsymbol{V}^{*}) \leq 1$.
Moreover, condition (\ref{Wt}) is satisfied only if $\mathsf{rank}(\boldsymbol{V}^{*}) \neq 0$, given that $\left ( d_0 \right )^2 \Gamma_{\min} > 0$.
Summarizing the above, we conclude that $\mathsf{rank}(\boldsymbol{V}^{*}) = 1$.
\end{proof}

\section{Proof of Proposition \ref{prop_conv}}
\label{app_conv}
\begin{proof}
Let $\boldsymbol{v}^i, \boldsymbol{W}^i, \boldsymbol{L}^i, \boldsymbol{F}^i$ denote the solution obtained in the $i$-th iteration of Algorithm 1, and $\Phi(\boldsymbol{v}^i, \boldsymbol{W}^i, \boldsymbol{L}^i, \boldsymbol{F}^i)$ represent the corresponding value of the objective function.
First, given $\boldsymbol{v}^i, \boldsymbol{W}^i, \boldsymbol{F}^i$, we solve the problem (\textbf{P2}) to get the task allocation of users to the ALAP $\boldsymbol{L}^{i+1}$.
Therefore, we obtain
\begin{equation}
\label{conv_1}
    \Phi(\boldsymbol{v}^i, \boldsymbol{W}^i, \boldsymbol{L}^i, \boldsymbol{F}^i) \geq \Phi(\boldsymbol{v}^i, \boldsymbol{W}^i, \boldsymbol{L}^{i+1}, \boldsymbol{F}^i).
\end{equation}
Then, we solve the problem (\textbf{P3}) to get the computation resource allocation of users and the ALAP $\boldsymbol{F}^{i+1}$ for given $\boldsymbol{v}^i, \boldsymbol{W}^i, \boldsymbol{L}^{i+1}$.
Hence, it follows that
\begin{equation}
\label{conv_2}
    \Phi(\boldsymbol{v}^i, \boldsymbol{W}^i, \boldsymbol{L}^{i+1}, \boldsymbol{F}^i) \geq \Phi(\boldsymbol{v}^i, \boldsymbol{W}^i, \boldsymbol{L}^{i+1}, \boldsymbol{F}^{i+1}).
\end{equation}
Accordingly, with $\boldsymbol{W}^{i}, \boldsymbol{L}^{i+1}, \boldsymbol{F}^{i+1}$ given, we can solve the problem (\textbf{P4A}) to get the beamforming vector at the ALAP transmitter $\boldsymbol{v}^{i+1}$.
Consequently, we have
\begin{equation}
\label{conv_3}
    \Phi(\boldsymbol{v}^i, \boldsymbol{W}^i, \boldsymbol{L}^{i+1}, \boldsymbol{F}^{i+1}) \geq \Phi(\boldsymbol{v}^{i+1}, \boldsymbol{W}^i, \boldsymbol{L}^{i+1}, \boldsymbol{F}^{i+1}).
\end{equation}
Finally, we solve the problem (\textbf{P5A}) to get the beamforming vectors at the ALAP receiver $\boldsymbol{W}^{i+1}$ for given $\boldsymbol{v}^{i+1}, \boldsymbol{L}^{i+1}, \boldsymbol{F}^{i+1}$.
Thus, we can obtain
\begin{equation}
\label{conv_4}
    \Phi(\boldsymbol{v}^{i+1}, \boldsymbol{W}^i, \boldsymbol{L}^{i+1}, \boldsymbol{F}^{i+1}) \geq \Phi(\boldsymbol{v}^{i+1}, \boldsymbol{W}^{i+1}, \boldsymbol{L}^{i+1}, \boldsymbol{F}^{i+1}).
\end{equation}
Combining (\ref{conv_1}), (\ref{conv_2}), (\ref{conv_3}) and (\ref{conv_4}), we can derive
\begin{equation}
    \Phi(\boldsymbol{v}^i, \boldsymbol{W}^i, \boldsymbol{L}^i, \boldsymbol{F}^i) \geq \Phi(\boldsymbol{v}^{i+1}, \boldsymbol{W}^{i+1}, \boldsymbol{L}^{i+1}, \boldsymbol{F}^{i+1}),
\end{equation}
which demonstrates that the value of the objective function does not increase over iterations.
Moreover, due to the finite range of the optimization variables, the total energy consumption remains bounded.
Hence, the convergence of Algorithm 1 is guaranteed.
\end{proof}

\end{appendices}

\bibliographystyle{IEEEtran}
\bibliography{reference}

\end{document}